\documentclass[preprint,amsmath,amssymb,pra]{revtex4-2}

\usepackage{graphicx}% Include figure files
\usepackage{epstopdf}
\usepackage{dcolumn}% Align table columns on decimal point
\usepackage{bm}% bold math
\usepackage{layouts}
\usepackage{ulem}
\usepackage{amsmath}
\usepackage{hyperref}% add hypertext capabilities
\hypersetup{
	colorlinks   = true,
	urlcolor     = blue,
	linkcolor    = blue,
	citecolor    = blue
}
\usepackage[dvipsnames]{xcolor}
\usepackage{float}
\usepackage{siunitx}

\begin{document}
	
	\title{Robust and compact single-lens crossed-beam optical dipole trap for Bose-Einstein condensation in microgravity}
	
	\author{J. S. Haase$^1$}
	\email{haase@iqo.uni-hannover.de}
	
	\author{A. Fieguth$^2$}
	\email{alexander.fieguth@dlr.de}
	
	\author{I. Bröckel$^2$}
	\author{J. Hamann$^1$}
	\author{J. Kruse$^2$}
	\author{C. Klempt$^{1,2}$}
	
	\affiliation{
		$^1$Institut für Quantenoptik, Leibniz Universität Hannover, Welfengarten 1, D-30167 Hannover, Germany \\
		$^2$Deutsches Zentrum für Luft- und Raumfahrt e.V. (DLR), Institut für Satellitengeodäsie und Inertialsensorik (DLR-SI), Callinstraße 30b, D-30167 Hannover, Germany
	}

\begin{abstract}
We present a novel concept for a compact and robust crossed-beam optical dipole trap (cODT) based on a single lens, designed for the efficient generation of Bose-Einstein condensates (BECs) under dynamic conditions. The system employs two independent two-dimensional acousto-optical deflectors (AODs) in combination with a single high-numerical-aperture lens to provide full three-dimensional control over the trap geometry, minimizing potential misalignments and ensuring long-term operational stability. By leveraging time-averaged potentials, rapid and efficient evaporative cooling sequences toward BECs are enabled. The functionality of the cODT under microgravity conditions has been successfully demonstrated in the Einstein-Elevator in Hannover, Germany, where the beam intersection was shown to remain stable throughout the microgravity phase of the flight. In addition, the system has been implemented in the sensor head of the INTENTAS project to verify BEC generation. Additional realization of one-, two-, and three-dimensional arrays of condensates through dynamic trap shaping was achieved. This versatile approach allows for advanced quantum sensing applications in mobile and space-based environments based on all-optical BECs.
\end{abstract}

\maketitle

\section{Introduction}
Interferometers based on laser cooled atoms can measure inertial forces with great accuracy~\cite{Peters1999, Gauguet2008, Stockton2011, Hu2013, Berg2015, Freier2016}. Bose-Einstein condensates (BECs) offer improvements of many systematic effects due to their large coherence lengths and the improved mode control in position and momentum. Their utilization is challenged by comparably long preparation times during evaporative cooling, which lead to sensor dead times and a reduced bandwidth. Atom chips, planar wire structures that magnetically trap the atoms nearby, have already been employed to reach short preparation times~\cite{Haensel2001,Farkas2014,Rudolph2015}, but restrict the optical access and deteriorate the quality of laser beams for the interferometric sequence. Furthermore,  they require magnetically sensitive states that must be transferred upon release from the trap. The generation of BECs in static optical traps~\cite{Barrett2001} is challenging, because size, trapping confinement, and trap depth cannot be chosen individually. While an approach with movable optical elements~\cite{Kinoshita2005} adds complexity and stability issues, time-averaged potentials from spatially modulated laser-beams~\cite{Henderson2009,Roy16} offer the fastest sizable BEC generation to date~\cite{Hetzel2025,Herbst2024}. However, such setups typically rely on two or more independent trapping laser beams from different directions, which make them susceptible to beam pointing instabilities. Although the progress in BEC generation in static traps is ongoing´, the atom flux is not on the same order of magnitude as in state of the art systems~\cite{Clement2009,Urvoy2019}. 

Here, we present a novel approach, where a single lens in combination with two two-dimensional acousto-optical deflectors (2D-AODs) provides three-dimensional control of the trap, while eliminating the main source of relative misalignment between the beams. The use of a single lens to generate a cODT is novel, although it is applied in a comparable way in quantum gas microscopes~\cite{Bakr2009, Sherson2010} and accordion lattice experiments~\cite{Fallani2005, Williams2008, Ville2017}. Comparable to~\cite{Condon2019, Pelluet24}, our setup is designed for the operation under microgravity while being able to withstand increased accelerations during launch and landing. It is thus also applicable for robust sensors in space or mobile applications. 
To demonstrate the functionality, we will showcase data and results based on its operation in the sensor head of the INTENTAS project~\cite{Anton24}.
The INTENTAS project exploits the described BEC generation to demonstrate entanglement-enhanced interferometry~\cite{Cassens2025} in a microgravity environment.
At the end of the article, we demonstrate how the experimental flexibility of painted potentials can be exploited for the simultaneous generation of multiple BECs, as it is desirable for spatially resolved sensing applications and the suppression of common-mode noise.

\section{Setup}

\begin{figure}[b]
    \centering
    \includegraphics[width=0.9\linewidth]{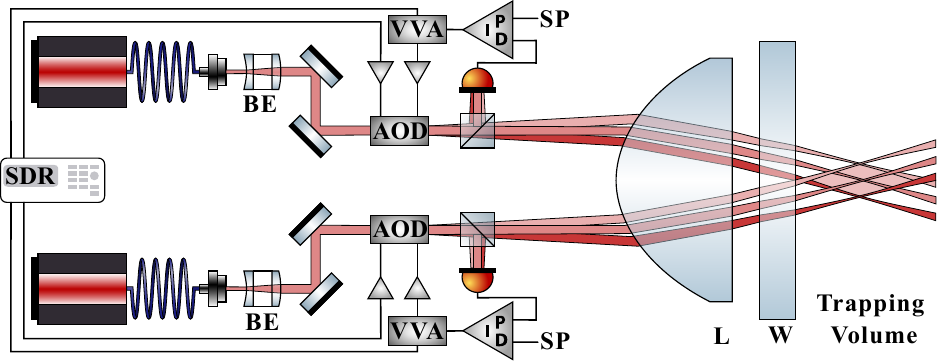}
    \label{fig:setup}

    \caption{System sketch of the novel crossed optical dipole trap. The key concept is the use of a single lens (L) with high numerical aperture (NA = 0.62) in combination with two two-dimensional acousto-optic deflectors (AODs), driven simultaneously by a software-defined radio (SDR) to generate time-averaged potentials. The two separate laser beams are expanded with beam expanders (BEs) to match the aperture of the AODs, after which their paths are intersected at 30$^\circ$ through the lens. By adjusting the AOD drive frequencies, the intersection can be maintained under displacements in all three dimensions. The beam intensities are stabilized via photodiodes feeding a PID loop that controls two voltage-variable attenuators (VVAs) with the set point (SP) as input. The additional window (W) represents the optical access to the INTENTAS vacuum chamber, where the trapping volume is indicated as red shaded area. A dichroic mirror used to separate the trapping and detection paths is omitted for clarity.}
    \label{fig:setup}
\end{figure}

The key component of the presented setup is an aspheric lens (Asphericon ALL75-60-P-U-780) as shown in Fig.~\ref{fig:setup}. The striking characteristic of this lens is its large numerical aperture of 0.62. The effective focal length of 60\,mm and the diameter of 75\,mm are chosen such that they are best suited for the application described in~\cite{Anton24}. Given the dimensions of the lens, two laser beams entering the lens in parallel at a distance of 30\,mm will cross at about 30$^\circ$ at the focal point.

The two beams can either be generated from a single laser source, and split before coupling into the system or can be generated separately. The latter approach was chosen for the application presented here in order to ensure enough power for the initial trapping. The two 1.064\,\textmu m laser beams are generated by two fiber-based lasers (NKT Koheras BASIK + BOOSTIK HP), each capable of providing a beam smaller than 0.65\,mm~(1/$e^2$ radius) with up to 15\,W of power at the fiber outcoupler. From there, each beam is sent into a beam expanding lens system consisting of a concave lens with f\,=\,-25\,mm (Thorlabs LC1054-C) and an achromatic doublet with f\,=\,75\,mm (Thorlabs AC127-075-C) setting the beam size to 1.95\,mm~(1/$e^2$ radius).
For larger beam sizes, the beam develops an increasing astigmatism, because the focused beam passes a 10\,mm vacuum window at an angle. The astigmatism therefore limits the achievable spot size, as horizontal and vertical focus points separate from each other.
We reach a focal spot with a diameter of 10.5\,\textmu m (1/$e^2$ radius) and a Rayleigh range of  320\,\textmu m. These beam parameters were measured in a separate implementation and agree with the results of an optics simulation (Zemax).

After the beam expansion, two mirrors (Thorlabs BB05-E03) per beam are used to couple each into a separate 2D-AOD (AAOptoelectronics DTSXY-400). This device is capable of steering the beam in two separate axes while maintaining an overall power efficiency of more than 75\,\% as measured in-situ. The AODs operate around a center frequency of 75\,MHz and provide a frequency range of $\pm$15\,MHz. This is equivalent to a deflection angle range of $\pm$\,1.4$^\circ$. The AODs also set a limit for the beam size given their aperture of 7.5\,$\times$\,7.5 \,mm$^2$. 
The output beams of the AODs are finally reflected under 90$^\circ$ by a dichroic mirror onto the aspheric lens. The dichroic mirror has a reflectance of 95\% at 1064 nm and a transmittance of 99\,\% at 780 nm, allowing to use the port also for fluorescence detection. The power at the focal point reaches a value of up to 10\,W per beam. \\
The rf signals to control all four channels of the AODs are initially generated by a software-defined radio (SDR) (Ettus USRP X410). The output signal passes through a voltage variable attenuator (VVA) (Minicircuits ZX73-2500-S+) in order to enable active intensity stabilization. For this purpose, the fraction of light which is transmitted by the dichroic mirror is collected by two photodiodes (Thorlabs PDA100A2), one for each beam, and the measured value is passed to a self-built analog PID controller (based on~\cite{Wendrich2010}) where an analog set point (SP) can be given. 
The SDR opens up the possibility to create time-averaged potentials by varying the frequency passed to the AODs. Thereby, the initial trap volume can be increased for efficient, fast evaporation~\cite{Roy16, Hetzel2025}. In our implementation, the relation between the actual displacement in the trap region and the AOD frequency is 86\,$\frac{\text{\textmu m}}{\text{MHz}}$ in vertical direction as measured by imaging displaced atoms with a calibrated camera. This is in agreement with a simulation using the actual geometry, which predicts a mean displacement of $88 \pm 5\,\frac{\text{\textmu m}}{\text{MHz}}$ in vertical direction and $92 \pm 4\,\frac{\text{\textmu m}}{\text{MHz}}$ in horizontal direction.

Now, a maximum trap volume can be defined by the maximally available deflection angles provided by the AODs.
The diamond-shaped area encapsulated in the plane of the beams is 27.2\,$\text{mm}^2$, as given by the maximal horizontal displacement of $\pm 1.38$\,mm and the relative beam angle.
An additional vertical  displacement of $\pm 1.32$\,mm yields a total volume of 71.8\,$\text{mm}^3$.
Given the available power, the trap depth in an unmodulated trap can be estimated to 11.9\,mK.
For an operational spatial modulation of $\pm$740\textmu m ($\pm8$\,MHz) in horizontal direction,  a maximum initial trap depth of 240\,\textmu K is achieved according to the simulated trap geometry.
In order to ensure the required robustness and sturdiness, while fulfilling requirements for weight and size, the whole beam path is encapsulated in a sturdy aluminum housing to minimize relative drifts of the two dipole trap beam paths.

\section{Performance}

\begin{figure}[b]

   \centering
   \includegraphics[width=\linewidth]{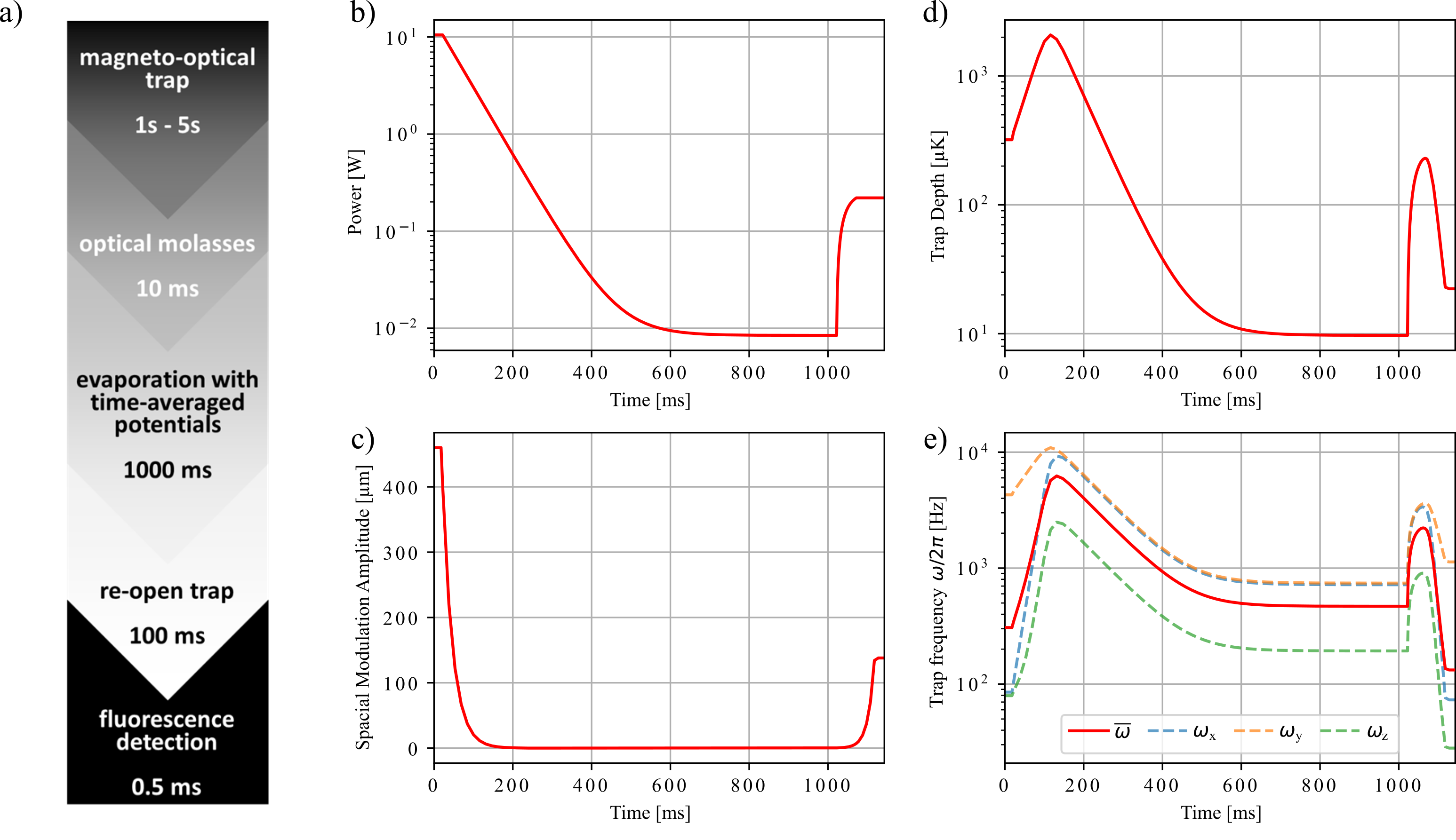}

   \caption{Experimental sequence and trap parameters of a BEC sequence. (a) Overview of the sequence with typical timings. During the evaporation, the beam power (b) and the spatial modulation amplitude (c) are varied to evaporatively cool the atoms to a BEC. These parameters allow calculating the optical dipole trap depth (d) and trap frequencies (e). The re-open phase at the end of the sequence creates aspect ratio to prove condensation via non-thermal expansion.} 
   \label{fig:TrapData}
\end{figure}

\begin{figure}[bp]
    \centering
        \centering
        \includegraphics[width=0.9\linewidth]{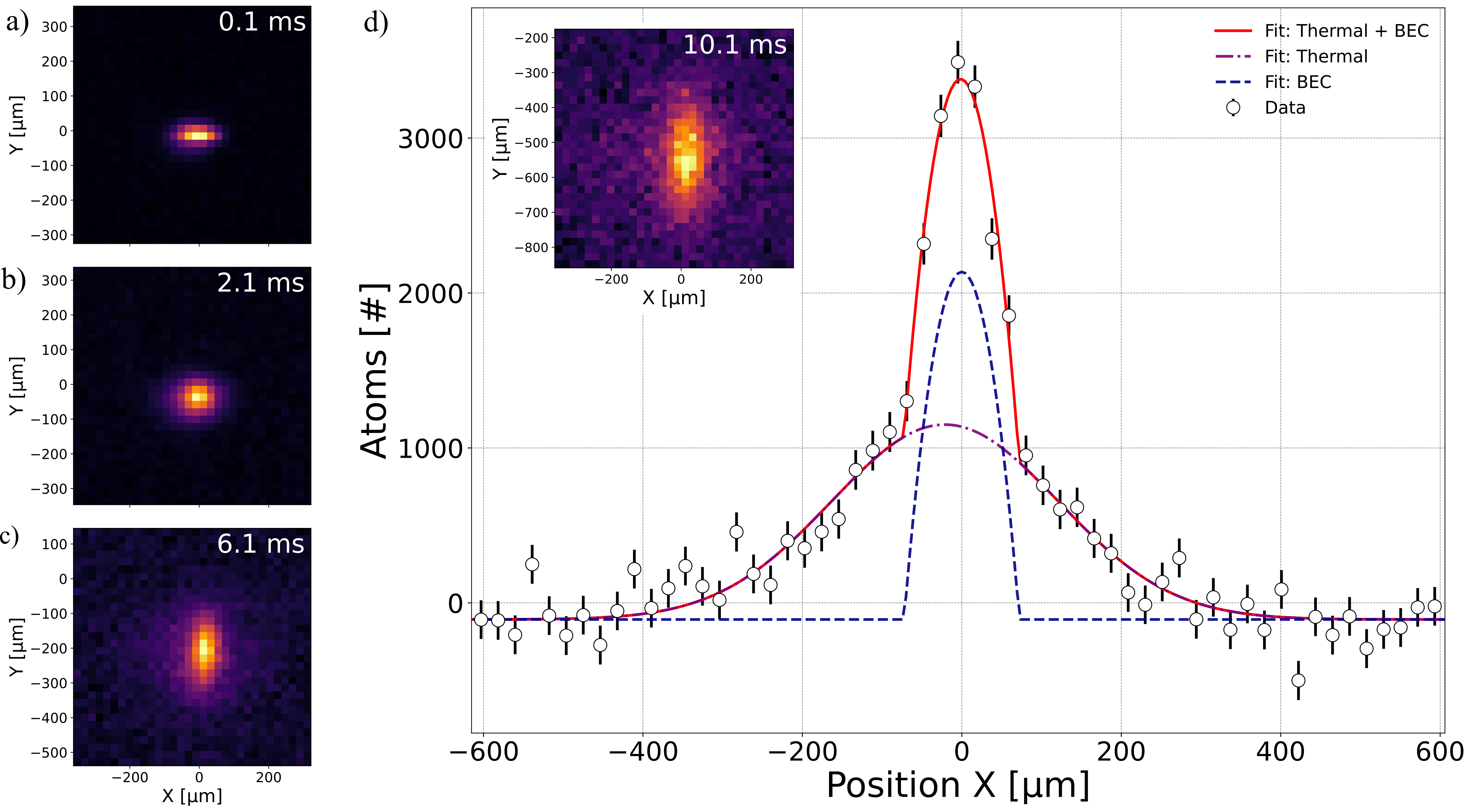}
    \caption{Non-thermal behavior of the atom cloud as proof of Bose-Einstein condensation. On the left side, three realizations with detection at different expansion times (0.1\,ms, 2.1\,ms, 6.1\,ms) are shown. The trap was initially elongated (relaxed) along the horizontal axis (a). During free fall, the thermal cloud expands isotropically while the BEC mainly expands along the axis with the highest trap frequency (i.e. the vertical direction). This leads to an inversion of the aspect ratio, which can be observed as a function of free-fall time (b, c). (d) For the data set at 10.1\,ms (shown in the inset), a projection onto the horizontal axis is shown. While a pure thermal expansion does not describe the data, a combination of a thermal cloud and a BEC (Thomas-Fermi model) is successfully describing the number of atoms as a function of position ($\chi^2_{\text{red}} = 1.4$).} 
    \label{fig:BEC_proof}
\end{figure}

We will now apply the presented dipole trap for the generation of BECs and test the setup's robustness towards a future microgravity application.
The sequence to achieve condensation consists of an initial MOT loaded with a (4$\text{--}$6)\,$\times$\,10$^{8}$ atoms followed by 10\,ms of optical molasses. This reduces the temperature of the atoms down to 30\text{--}50\,\textmu K. The dipole trap is filled during the molasses phase with 2\,$\times$\,10$^{6}$ atoms at a temperature of 20\,\textmu K, being consistent with the truncation parameter given in~\cite{Roy16}. With the simulated trap frequency the initial phase space density is 1.15\,$\times$\,10$^{-3}$, comparable to~\cite{Barrett2001}. An evaporation sequence motivated by~\cite{Hetzel2025} follows. The sequence includes an exponentially decreasing intensity ramp, where the power is reduced from 10\,W to 40\,mW within 1\,s. Simultaneously, the spatial modulation amplitude is ramped down from $\pm$460\,\textmu m (5\,MHz to 0\,MHz) exponentially within the first 200\,ms. In order to stop the evaporation sequence, the intensity of the final configuration is slightly increased. To test if condensation has been achieved, the spatial modulation amplitude is increased in two dimensions by $\pm$140\,\textmu m (1.5\,MHz). By changing the intensity, the trap depth can be increased while the trap frequency is decreased before release (and after evaporation) in those two directions. This allows to observe the characteristic behavior of the BEC, i.e. an aspect-ratio inversion due to the discrepancy in trap frequencies rather than the isotropical uniform expansion expected from a thermal cloud (Fig.~\ref{fig:BEC_proof}).
The time evolution of the trap parameters is visualized in Fig.~\ref{fig:TrapData}. An additional analysis of the density distribution in the horizontal direction using a combined fit of a thermal cloud and a BEC after a ballistic expansion of 10.1\,ms reveals that partial condensation has been successfully achieved. 
Notably, the power of the final trap has been chosen to observe the mixture of the thermal cloud and the BEC. A decrease in power removes the detectable thermal fraction and a BEC with $\mathcal{O}$(10$^{\text{4}}$) atoms remains. The evaporation efficiency is calculated to be $\gamma = 2.7$ for $10^4$ atoms in the BEC \cite{Roy16}. 
The full sequence has been performed within 2\,s, which fulfills the requirement set by the available free-fall time of 4\,s in the Einstein-Elevator in Hannover~\cite{Lotz22}.

\begin{figure}[t!!]
    \centering
    \includegraphics[width=\linewidth]{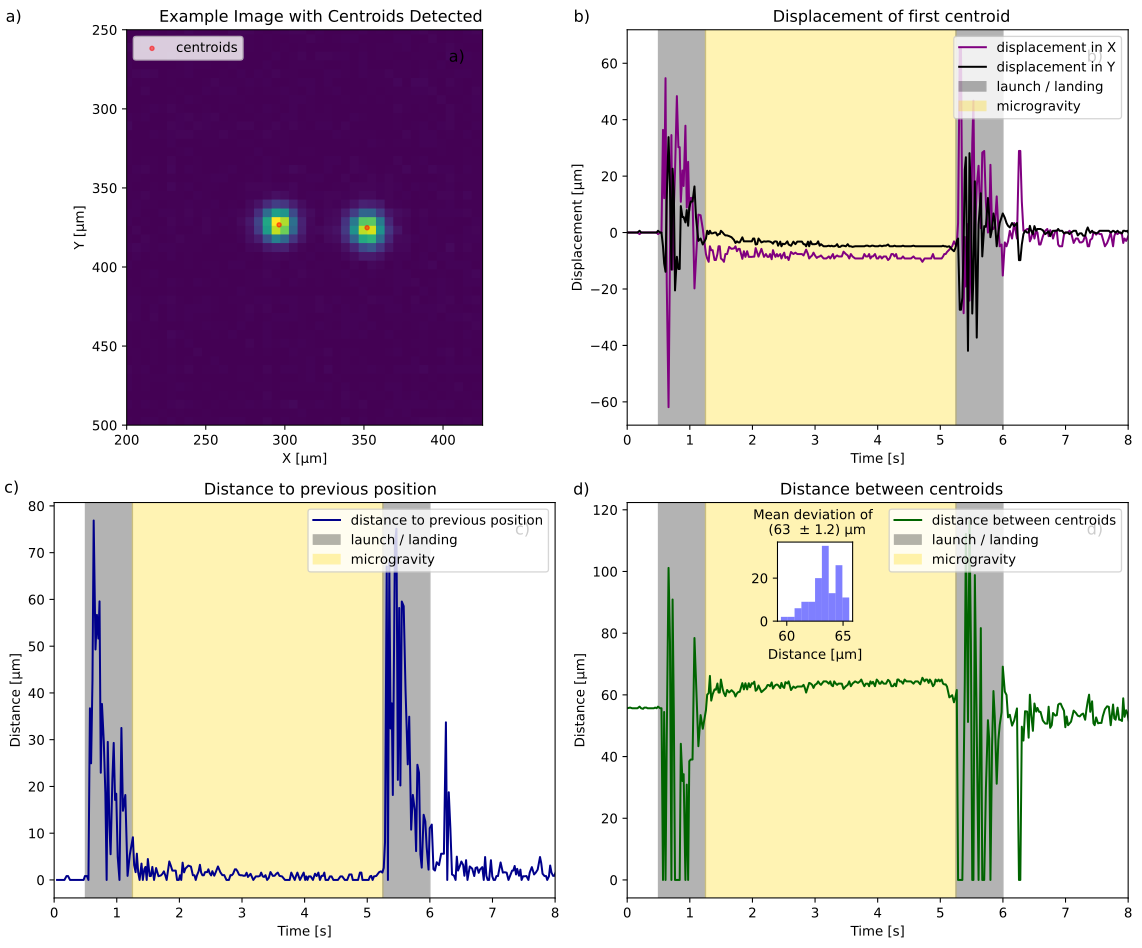}
    \caption{Exemplary results of a single test flight of the dipole trap setup in the Einstein-Elevator. (a) Camera image with the two laser spots recorded as well as the calculated centroids used for the subsequent positional analysis. (b) Displacement in the x (purple) and y (black) axis for one of the laser spots. (c) The Euclidean distance of a laser spot between consecutive images (blue). (d) Euclidean distance between the two spots in the same image (green). The launch and landing are indicated in the analyzed data (gray shaded) as well as the time in microgravity (gold shaded) which is relevant for any assumption of the capability of the system for the atomic sensors. The inset shows the spread of the distance during microgravity once the central position has settled.}
    \label{fig:test_flight}
\end{figure}

In order to ensure its functionality in microgravity, especially regarding the critical relative alignment, a dedicated test setup using the dipole trap hardware has been built and operated in the Einstein-Elevator. The setup consisted of an early version of the dipole trap structure as described previously with a few differences and shortcomings in comparison to the final realization. 
A single 1064\,nm laser source (NP Photonics The Rock OEM Module) was used and its signal was split by a fiber beam splitter (Thorlabs PN1064R5A1) before being coupled into both AOD paths, with a beam size of 2\,mm~(1/$e^2$ radius) (Thorlabs TC18APC-1064). While an identical lens as in the final setup has been used, it was not attached to the experimental chamber but was mounted on two 105\,mm long posts. At the focal position of the lens a camera (WinCamD-LCM, DataRay) has been mounted on a post as well. The camera was used to monitor the spot position of each laser beam in the horizontal and vertical direction with 24\,fps. 
The setup then has been installed in the Einstein-Elevator and a test flight under comparable conditions to a scientific run has been performed. The position of the beam spots, which have been intentionally placed about 60\,\textmu m apart in the horizontal axis of the camera, have been recorded before launch, during the entire flight period and after landing. The results of an exemplary flight can be seen in Fig~\ref{fig:test_flight}. For the subsequent analysis, a threshold mask was applied to the images and shape-finding using the skimage package (Python)~\cite{Van2014} was used to determine the centroids of each spot.

In order to ensure functionality of the cODT during an experimental run one needs to verify that the position of the laser beams, and therefore the intersection of the two laser beams, does not vary uncontrollably compared to the waist of the respective laser beams during the microgravity operation (an AC displacement type of misalignment). 
Additionally, if there is a shift after launch or  landing, it needs to be within the range of corrections available by adjusting the AOD central frequencies (a DC displacement type of misalignment). 
We observe a noticeable DC displacement from the original position, mainly along a diagonal axis, of both beams during acceleration and deceleration. However, the data shows that it does not exceed 75\,\textmu m in any particular axis (Fig.~\ref{fig:test_flight} (b)). If launch and landing are excluded, however, the displacement settles to less than 12\,\textmu m during the microgravity phase and recovers after landing to within a few \textmu m from the starting position. The AC displacement, i.e the variation from one image to the next (Fig.~\ref{fig:test_flight} (c)) has been measured to be less than 3\,\textmu m within the resolution of the camera. Since the mounting of lens and camera with posts is much more susceptible to movements than the final fixation to the experimental chamber, we expect some extraneous displacements, which will not be present in the final setup. One way to exclude common-mode movements, which would not impact the intersection of the two beams, and thus not change the trap depth, is to measure the relative distance of the two laser spots during microgravity. In  Fig.~\ref{fig:test_flight} (d), an analysis of the relative distance change of the two spots reveals a deviation of about 5\,\textmu m to the initial position, with a standard deviation of 1.2\,\textmu m. The latter is even smaller for the inner three seconds of the free fall, where the main operation of the cODT will take place.
In order to test the result of such deviations on the BEC generation performance, we have performed lab experiments with an intentional misalignment of the beams.
We find that the total number of atoms is stable to within 30\% for a relative misalignment of the two beams for up to $\pm$8\,\textmu m.
This compares favorably to the 5\,\textmu m displacement from ground to micro-gravity operation as an upper limit for the expected relative fluctuations.
We thus conclude that the observed mechanical stability is sufficient for the creation of all-optical BECs under microgravity. 

\section{Application}

\begin{figure}[b]
    \centering
    \includegraphics[width=\linewidth]{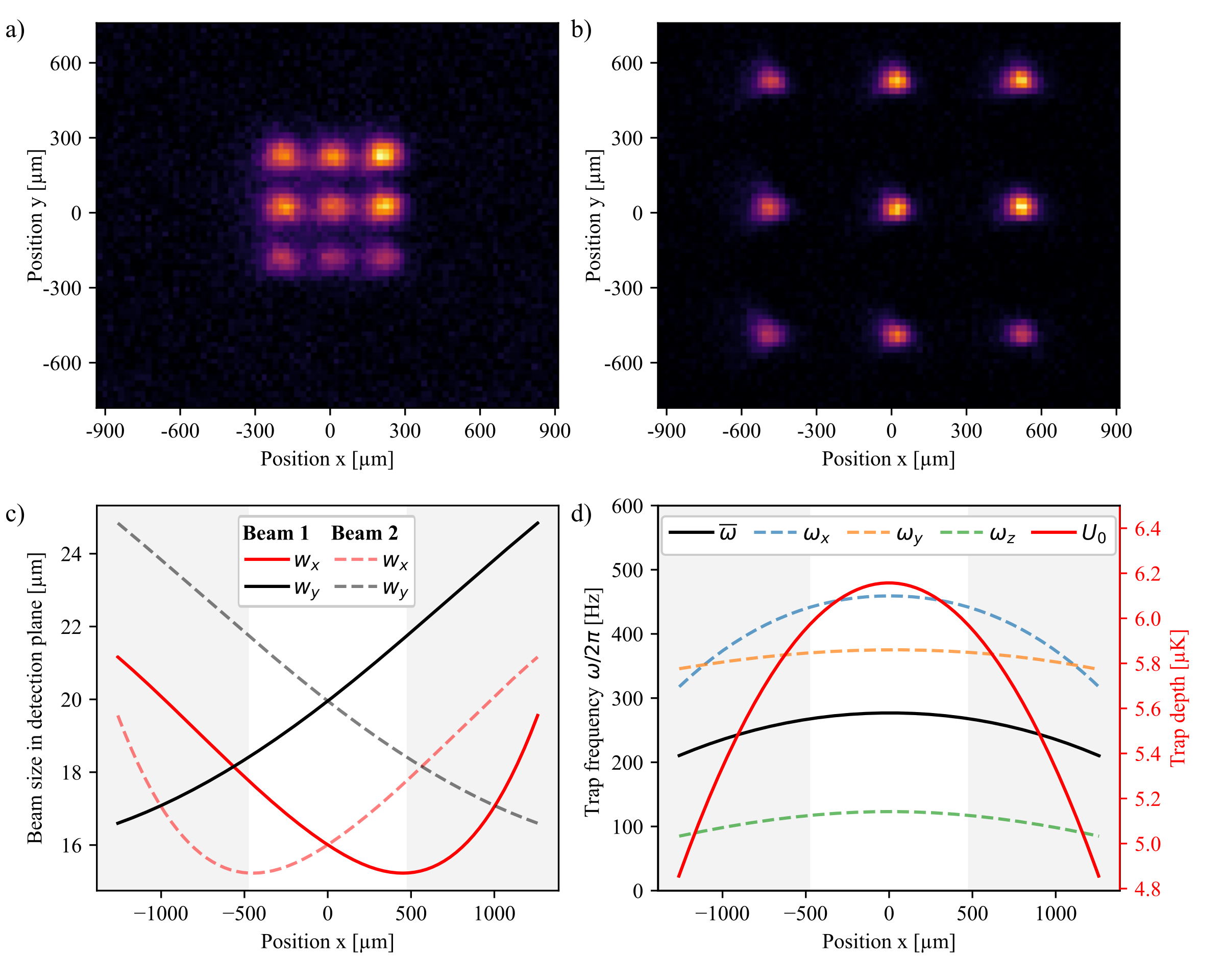}
    \caption{3\,$\times$\,3 grid of atom clouds with 190\,\textmu m (a) and 480\,\textmu m (b)  spacing. Each cloud consists of 5,000 to 10,000 atoms. c) Beam sizes at the intersection of the two beams in the detection plane simulated at different $x$ positions. Corresponding changes of trap frequencies and trap depths are shown in d).}
    \label{fig:bec}
\end{figure}

The full three-dimensional control of the setup allows creating and designing time-averaged potentials, which extend its reach beyond single Bose-Einstein condensation~\cite{Ryu2013,Ryu2014, Eliasson2019, Stolzenberg2025}. For example, the generation of one-, two- and three dimensional arrays of BECs is possible. 
In order to create a one-dimensional array in the horizontal direction, the time-averaged potential is adiabatically transformed from a harmonic potential into a multi-well potential, distributing the atomic ensembles into separated potential-wells. Each of these wells is then further reduced to induce evaporation in order to achieve Bose-Einstein condensation.
For the vertical direction, one already starts with a time-averaged potential in spatially separated traps vertically spaced by the respective AODs. This means that the modulation is still applied in the horizontal direction, but the multiple intersections in vertical direction allow for direct trapping of multiple atomic ensembles, which are subsequently cooled down towards Bose-Einstein condensation.
In vertical direction, independent traps are superior to splitting a single trap, because gravity would lead to an asymmetric splitting ratio.
Combining both methods, it is possible to create two-dimensional arrays. The 3$\times$3 array in Fig.~\ref{fig:bec}(a) is created with a spacing of 190\,\textmu m between neighboring clouds. If a frequency ramp is applied at the AODs, the individual clouds can be transported after splitting and cooling. Exemplary, Fig.~\ref{fig:bec}(b) shows a 3$\times$3 array with a 480\,\textmu m spacing. 

Simulations show that the beam sizes change on the order of ±10\,\% for the 480\,\textmu m grid. Because the beam sizes of the two individual beams change in opposite direction (Fig. \ref{fig:bec}(c)), the effects on trap depth and trap frequency compensate each other (Fig. \ref{fig:bec}(d)). The mean trap frequency and the trap depth deviate by less than 3\,\%, the individual trap frequencies by less than 5\,\%. In principle, these deviations can be suppressed by adjusting the beam power in the individual grid sites. Changing the intensity to hold the trap depth constant will also reduce the deviation of the trap frequencies to below 2\,\%.
Extending the analysis to the full accessible modulation amplitude of $\pm 1.38$~mm yields a beam size deviation of less than 30\%. The mean trap frequency and the trap depth vary by 25\,\%, because the counteraction is inferior at the outer parts of the grid. Again, individual power adjustments can mitigate the inhomogeneity.
Also a transport out of the plane is possible, enabling three-dimensional positioning of the atoms. A transport out of plane of 330\,\,\textmu m and back into plane was performed in the system with an atom loss below 20\%. Due to restrictions of the detection system, the clouds had to be moved back to the focal plane to be analyzed, so the mode of operation can be further optimized to reduce atom loss and enhance usability. Further investigation of the three-dimensional transport is foreseen with a dedicated second detection system.

\newpage
\clearpage
\section{Conclusion}

We have presented a setup for a compact, robust and versatile cODT using a novel single-lens approach. This mitigates re-alignment challenges in dynamic environments and is less expansive in the required space for the apparatus. The combination with time-averaged potentials relying on two-dimensional AODs and high-power fiber lasers offers full three-dimensional manipulation of atom clouds and allows us to create single BECs as well as \mbox{one-,} two- and three-dimensional arrays of condensates. A further exploration of more exotic potentials than the presented harmonic potentials painted within the sizable volume of about 70\,mm$^3$ could open up new avenues for the manipulation of atoms. One such example would be the implementation of box-shaped potentials, which in principle should allow for larger atom clouds at uniform trap depth.  
A proof-of-concept using a test setup for multiple launches in the Einstein-Elevator in Hannover has shown its capability of performing in a microgravity environment. With deviations of the relative beam positions on the order of 1\,\textmu m, the intersection will be maintained during flight operation. This concludes that the cODT is ready to be deployed on atoms for BEC creation in microgravity as planned within the INTENTAS project. However, the application is not limited to the specific project, but extends to applications, that have a demand for a compact, robust cODT with three dimensional control over its intersection point without the necessity of magnetic fields in direct proximity of the atom clouds.  

\section*{List of Abbreviations}

\begin{tabular}{ll}
\textbf{Abbreviation} & \textbf{Full Term} \\
AOD        & Acousto-Optical Deflector \\
2D-AOD     & Two-Dimensional Acousto-Optical Deflector \\
BEC        & Bose-Einstein Condensate \\
cODT       & Crossed Optical Dipole Trap \\
MOT        & Magneto-Optical Trap \\
PID        & Proportional–Integral–Derivative Controller \\
SDR        & Software-Defined Radio \\
VVA        & Voltage-Variable Attenuator \\
\end{tabular}

\section{Declarations}

\subsection*{Ethics approval and consent to participate}
Not applicable

\subsection*{Consent for publication}
Not applicable

\subsection*{Availability of data and materials}
Data are available upon reasonable request.

\subsection*{Competing interests}
All authors declare that they have no competing interests.

\subsection*{Authors contributions}
A.F. drafted and edited the manuscript. J.S.H contributed to the editing, writing and figures.  C.K., J.K. contributed to the writing. J.S.H conceived the original concept and J.S.H, J.H., C.K. and J.K. were involved in the refinement of the setup. J.S.H and A.F. performed the measurements and data analysis. All authors reviewed the manuscript.

\subsection*{Funding}
This work is supported by the German Space Agency (DLR) with funds provided by the BMWK
under Grant No. 50WM2174. Supporting work was also contributed by the Deutsche Forschungsgemeinschaft (DFG) under
Germany’s Excellence Strategy within the Cluster of Excellence QuantumFrontiers (EXC
2123, Project ID 390837967).

\subsection*{Acknowledgements}

We would like to thank Dominik Köster for the initial work on this setup. Also, we would like to thank Dorothee Tell, Martin Quensen and Mareike Hetzel for their support regarding  time-averaged potentials. Furthermore, would like to thank Alexander Heidt, Sebastian Lazar, Moritz Möbius and Christoph Lotz for their assistance during installation of the test setup and operation of the Einstein-Elevator.

\bibliography{dipole2}

\end{document}